\documentclass[a4paper]{jpconf}
\pdfoutput=1

\usepackage{graphicx}
\usepackage{amsmath}
\usepackage{amsfonts}
\usepackage{amssymb}
\usepackage{hyperref}
\usepackage{slashed}

 \def\be   {\begin{equation}}   \def\ee   {\end{equation}}
 \def\ba   {\begin{array}}      \def\ea   {\end{array}}
 \def\bea  {\begin{eqnarray}}   \def\eea  {\end{eqnarray}}
 \def\bean {\begin{eqnarray*}}  \def\eean {\end{eqnarray*}}

\newcommand{\TeV}{{\rm \,TeV}}

\begin{document}

\title{Electroweak lights from Dark Matter annihilations}

\author{Andrea De Simone}

\address{Institut de Th\'eorie des Ph\'enom\`enes Physiques,
 \'Ecole Polytechnique F\'ed\'erale de Lausanne, CH-1015 Lausanne,
Switzerland}

\ead{andrea.desimone@epfl.ch}

\begin{abstract}
The energy spectra of Standard Model particles originated from Dark Matter annihilations
can be significantly altered by the inclusion of electroweak gauge boson radiation from the final
state.  A situation where this effect is particularly important is when a Majorana Dark Matter particle annihilates into two light fermions. This process is in $p$-wave and hence suppressed by the small value of the relative velocity of the annihilating particles. The inclusion of electroweak radiation eludes this suppression and opens up a potentially sizeable $s$-wave contribution to the annihilation cross section. I will 
discuss the impact of this effect on the fluxes of stable particles resulting from the Dark Matter annihilations, 
which are relevant for Dark Matter indirect searches. 
\end{abstract}

\section{Introduction}

The fluxes of stable Standard Model particles that originate from the annihilation 
(or decay) of Dark Matter (DM) in the galactic halo 
are the primary observable for DM indirect searches.
The radiation of ElectroWeak (EW) gauge bosons from the final state of the annihilation
process  turns out to have a great influence on
the energy spectra of stable particles and hence on the predictions for fluxes to be measured at Earth 
\cite{paper1, paperFSR}  (see also Ref.~\cite{ibarra}).
In particular, there are three situations where the effect of including the EW corrections is especially important: 
\begin{enumerate}
\item  when the  low-energy regions of the spectra, which are largely populated by the decay products of the emitted gauge bosons, are the ones contributing the most to the 
observed fluxes of stable particles;
\item  when some particle species are absent if EW corrections are not included, e.g.~antiprotons from 
$W/Z$ decays in an otherwise purely leptonic channel;
\item 
when the $2\to3$ annihilation cross section, with soft gauge boson emission, is comparable or even dominant with respect to the $2\to 2$ cross section.
\end{enumerate}
In this talk I will focus on case (iii).
A possible situation realizing it occurs when one considers the DM as a gauge-singlet 
Majorana particle $\chi$ 
of mass $M_\chi$ annihilating into light fermions $f$ of mass $m_f\ll M_\chi$; 
it is well known
that in this case the $\chi\chi \to f\bar f$ cross section  is  suppressed. 
In fact,  one can perform the usual expansion of the cross section 
$v \sigma  =a+b\,v^{2}+\mathcal{O}(v^4)$,
where $v\sim 10^{-3}$ is the relative velocity (in units of $c$) of the DM particles in our galaxy today.
The $v$-independent term, corresponding to the annihilation of particles in $s$-wave, is constrained by helicity arguments to be proportional to  $(m_f/M_\chi)^2$, and hence very small for light final state fermions.
The second term, corresponding to $p$-wave annihilation, suffers from the $v^2$ suppression.
For a DM particle singlet under the SM gauge group,  the radiation of EW gauge bosons from the final state and
from the internal propagator of the annihilation process 
eludes the suppressions
and opens up a potentially sizeable $s$-wave contribution to the cross section
(see Ref.~\cite{bergstrom} for the case of photon radiation
and Ref.~\cite{drees} for gluon radiation).

There is actually another situation realizing the condition 3 above, 
where we expect therefore the EW corrections to have a great impact \cite{paperISR}. 
Relaxing the hypothesis that $\chi$ is a gauge singlet, 
and considering the possibility  that the DM is part of a multiplet  charged under the EW interactions,
then the DM annihilates predominantly in $s$-wave into $W^+W^-$, if kinematically allowed.
However, now even the initial state of the channel $\chi\chi\to f\bar f$  can radiate a gauge boson; 
this process also lifts the helicity suppression and contributes to the  $s$-wave cross section,
becoming competitive with the di-boson channel.
We refer the reader to Ref.~\cite{paperISR} for more details.
%

\section{The model}
\label{sec:generalsettings}

In this section, I present the (toy) model I will use to describe more concretely the relevance
of the EW corrections in DM annihilations.
Let us add to the particle content of the SM a
 Majorana spinor $\chi$ with mass $M_\chi$,  singlet under the SM gauge
 group and playing the role of DM,
and a scalar $SU(2)$-doublet $S=(\eta^+, \eta^0)^T$, with mass $M_S>M_\chi$.
The field $S$ provides the interactions
of the DM with the generic fermion of the SM, described by the left-handed doublet
 $L=(f_1, f_2)$. In fact, the total  Lagrangian of the model is
\be
{\cal L}={\cal L}_{\rm SM}
+{1\over 2}{\bar{\chi}(i\slashed{\partial}-M_\chi)\chi}
+(D_\mu S)^\dagger (D^\mu S)-M_S^2S^\dagger S
+[y_L \bar{\chi}(Li\sigma_2 S)+{\rm h.c.}]\,.
\label{eq:Lagrangian}
\ee
The parameter $M_S$ will be traded for the ratio $r\equiv M_S^2/M_\chi^2$.
The stability of the DM can be achieved e.g. by endowing $\chi$ and $S$ with odd parity
under a $Z_2$ symmetry, while the rest of the SM spectrum is even.
In a  supersymmetric framework,  one  recognizes the same interactions of a Bino
with fermions and their supersymmetric scalar partners.
We will restrict our attention to the
massless limit $m_{f_1}=m_{f_2}=0$. 
The generalization of our calculations to non-zero fermion masses will not be addressed here.

\section{Helicity suppression and its removal}

\begin{figure}[t]
\includegraphics[width=3.8cm, height=2.7cm]{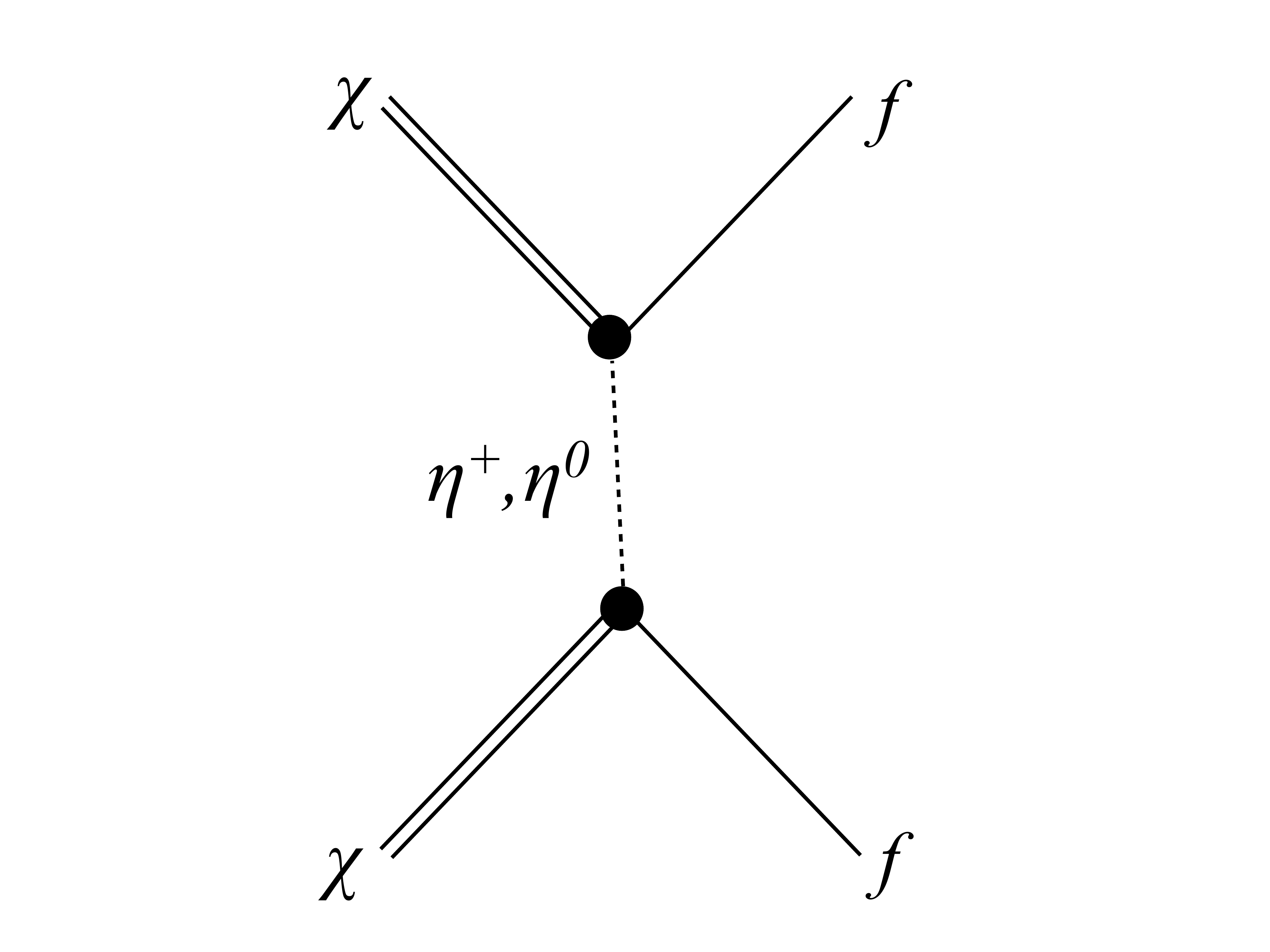}
\includegraphics[width=4cm, height=3cm]{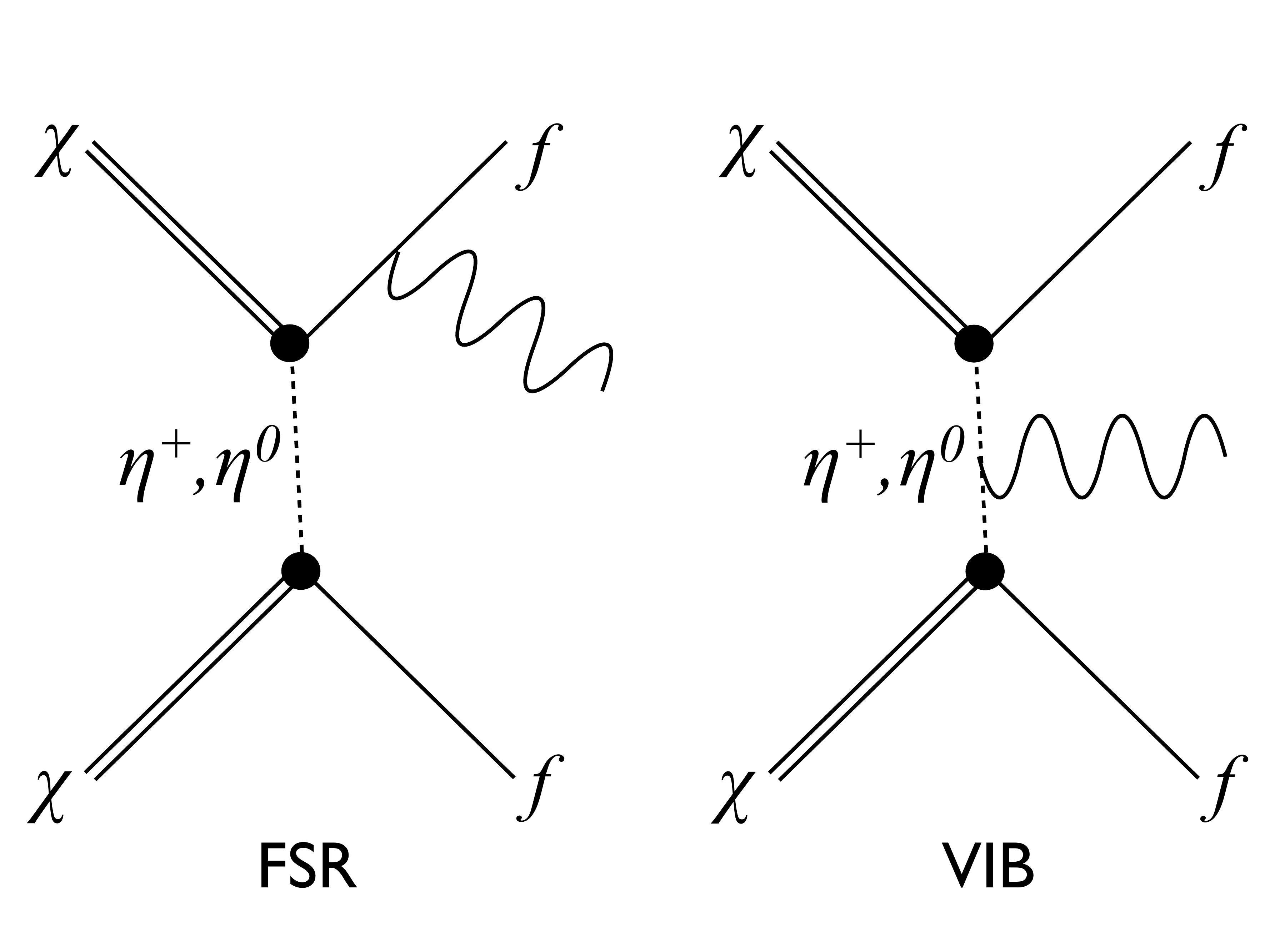}
\hspace{0.5cm}
\begin{minipage}[b]{7cm}
\caption{
\label{fig:diagrams}
Diagrams for 2-body (left panel) and 3-body (right panel) DM annihilations.}
\label{diagrams}
\end{minipage}
\end{figure}

\noindent
As it is well-known, the first non-zero contribution to the tree-level cross section of 
the annihilation of a Majorana fermion into a pair of massless left-handed fermions
(left panel of  Fig.~\ref{diagrams})
is velocity dependent, and hence suppressed, 
and is approximately  given by (see Ref.~\cite{paperFSR} for the exact expression)
\be
v\sigma(\chi\chi\to f\bar f)\sim {1\over M_\chi^2}{v^2\over r^2}\,.
\label{estimate2body}
\ee
The inclusion of higher-order processes with the emission of soft weak gauge bosons
 evades the helicity suppression and turns on an unsuppressed $s$-wave contribution to
 the DM annihilation cross section.
Let us consider,  for definiteness,  the 3-body process 
with the emission of a $Z$ boson
$\chi\chi\to \bar{f}_Lf_LZ$ (right panel of  Fig.~\ref{diagrams}).
Following Ref.~\cite{bergstrom}, we will call ``FSR'' (final state radiation) the processes where a gauge boson is radiated
from an external leg, while we refer to the emission from internal virtual particles as ``VIB'' (virtual internal bremsstrahlung). I am not going to write down any detailed expression for the amplitude,
and I refer the interested reader to Ref.~\cite{paperFSR} where the exact analytical results can be 
found; here, I  would just like to discuss
the structure of the amplitude, where the various contributions can be organized as follows
\be
{\cal M}\sim {1\over M_\chi}{\cal O}(v)\left [ \left.{\cal O}\left({1\over r}\right)\right\vert_{\rm FSR}
+ \left. {\cal O}\left({1\over r^2}\right)\right\vert_{\rm FSR} \right]+
{1\over M_\chi}
\left[  \left.{\cal O}\left({1\over r^2}\right)\right\vert_{\rm VIB} + \left.{\cal O}\left({1\over r^2}\right)\right\vert_{\rm FSR} \right]\,.
\label{Mestimate}
\ee
At this point we can learn an important lesson:
the opening of the $s$-wave originates from diagrams of both FSR and VIB type, at  
$\mathcal{O}(1/r^2)$ in the amplitude; limiting the expansion up to $\mathcal{O}(1/r)$ in the amplitude would cause
the process to stay in the $p$-wave.
An order-of-magnitude estimate  for the 3-body cross section,
showing the leading dependence on the expansion parameters,
can be obtained straightforwardly
\be
v\sigma(\chi\chi\to f\bar f Z)\sim \frac{\alpha_W}{M_{\chi}^2}\left[
 {\cal O}\left({v^2\over r^2}\right)+
{\cal O}\left({v^2\over r^3}\right)+
{\cal O}\left({1\over r^4}\right)
\right]\, ,
\label{estimate3body}
\ee
where  the weak coupling  $\alpha_W=g^2/(4\pi)$
for the gauge boson emission has been restored.
The estimates in Eqs.~(\ref{estimate2body}) and (\ref{estimate3body}) allow to gather an understanding in simple terms  of the
situation we are studying.
While the 2-body annihilation cross section behaves like $v^2/r^2$, the 3-body  FSR and VIB diagrams give rise
to both $s$-wave and $p$-wave terms;
the latter cannot compete with the 2-body cross section because of the
extra $\alpha_W$ factor; however, the former, free from the $v^2$ suppression, can overcome the $2 \to 2$
cross section if $r$ is not too large.

Because of the importance of this point, and being the distinction between FSR and VIB not able to disentangle clearly the $s$-wave contribution from the $p$-wave one, let us introduce now a specific notation. Having in mind an expansion in powers
of $1/r$ in the amplitude -- as sketched in Eq.~(\ref{Mestimate}) -- we will call ``leading order'' (LO) the lowest order
term $\mathcal{O}(1/r)$ in this expansion, which originates from lowest order FSR-type diagrams.
As shown above, in the LO approximation the annihilation cross section proceeds through $p$-wave.  Only 
higher order terms are able to remove the helicity suppression.

\section{Energy spectra of final stable particles}
\label{sec:spectra}

Let us now look at the energy spectra of stable particles 
produced by DM annihilations, with the inclusion of EW bremsstrahlung,
focusing in particular on positrons, antiprotons, photons  and neutrinos.
For simplicity, I consider  only the case  where $L=( \nu_{e\,L}, e_L)$ and 
the  primary annihilation channels for $\chi\chi\to I$, including EW bremsstrahlung, are
\begin{equation}
I= \{ e^+_L e^-_L,\, {\nu}_{e\,L}\bar {\nu}_{e\,L},\,
e^+_L e^-_L\gamma, \, e^+_L {\nu}_{e\,L} W^-,\, e^-_L\bar{\nu}_{e\,L}\, W^+, \,
e^+_L e^-_L Z, \, {\nu}_{e\,L}\bar{\nu}_{e\,L} Z
\}
 \,.
  \label{eq:channels}
\end{equation}
Using the  techniques described in Ref.~\cite{paperFSR} one is able
to  combine the analytical description of the primary annihilation 
channels with the numerical simulations for subsequent hadronization and decay 
and it is possible to extract the energy distributions of  each stable particle $f$
\be
\label{eq:norm}
\frac{d{\cal N}_f}{d \ln x}\equiv {1\over \sigma_0}{d\sigma(\chi\chi\to f+\textrm{anything})\over d\ln x}
 \,,\qquad f=\{ e^+, e^-, \gamma, \nu,\bar\nu, p,\bar p\}\,,
\ee
where $x\equiv E_{\rm kinetic}^{(f)}/M_{\chi}$, $E_{\rm kinetic}^{(f)}$ is the kinetic energy of the particle $f$ (the difference between total and kinetic energies is obviously relevant only for the (anti)protons).
The normalization is chosen to be the tree-level cross section of the 2-body processes
$\sigma_{0}=\sigma_{\rm tree}(\chi\chi\to e^+_Le^-_L)+\sigma_{\rm tree}(\chi\chi\to \nu_{e\,L}\bar{\nu}_{e\,L})$.
\begin{figure}[t]
\includegraphics[scale=0.6]{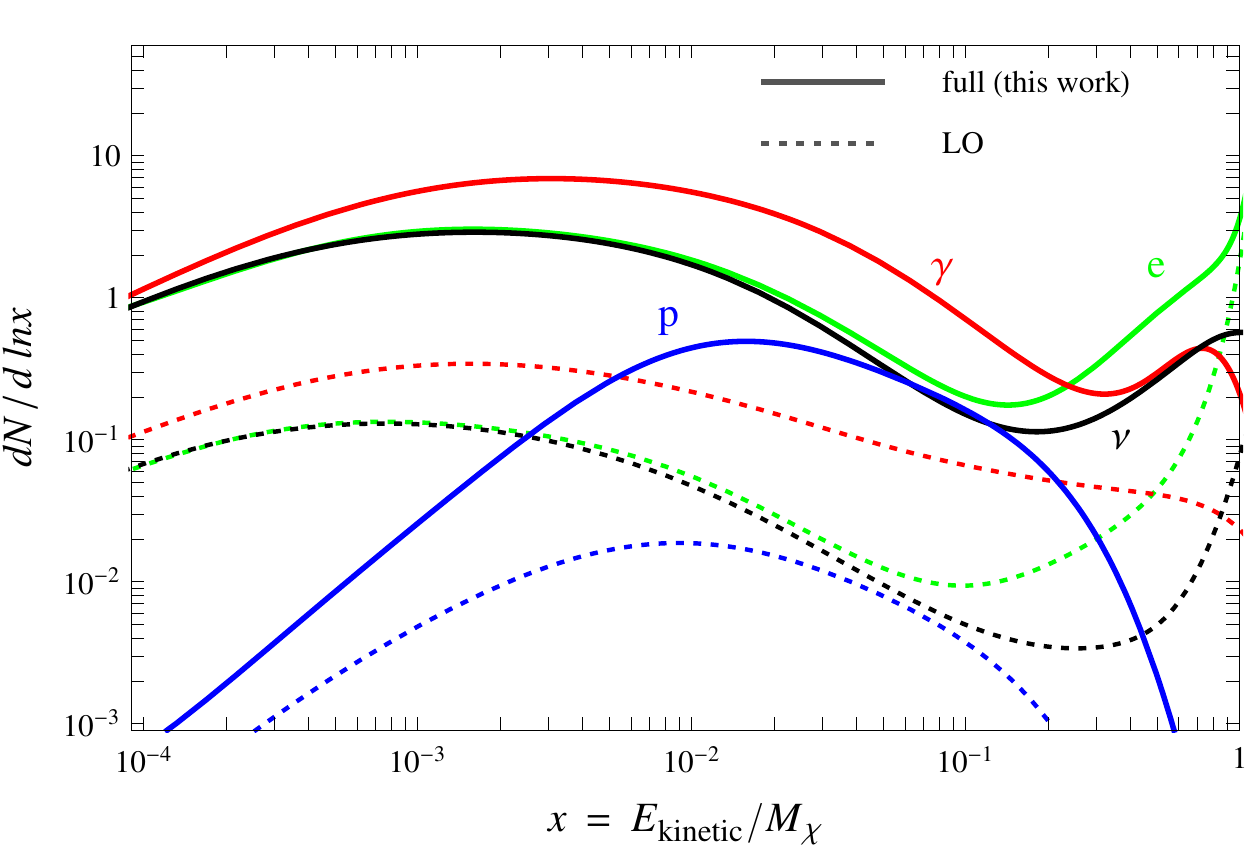}\hspace{2pc}%
\begin{minipage}[b]{7.4cm}
\caption{
\label{fig:spectra}
The spectra ${d{\cal N}_f}/{d \ln x}$, as defined in Eq.~(\ref{eq:norm}), for
  $e^+$ (green), $\gamma$ (red), $\nu=(\nu_e+\nu_\mu+\nu_\tau)/3$ (black)
 and $\bar p$ (blue),  from the annihilation $\chi\chi\to e_L^+ e_L^-, \nu_{e\, L}\bar{\nu}_{e\,L}$
 with the corresponding weak boson emission corrections. for the case  
 $M_\chi=1 \TeV, M_S=4 \TeV, v=10^{-3}$ (solid lines).
 For comparison, we show  the spectra  (dashed lines) in the LO approximation.}
\end{minipage}
\end{figure}
The plot  in Figure \ref{fig:spectra} shows  the resulting ${d{\cal N}_f}/{d \ln x}$  for $e^+, \gamma, \nu=(\nu_e+\nu_\mu+\nu_\tau)/3, 
\bar p$ for a specific, but representative, choice of parameters:
$M_\chi=1 \TeV$, $M_S=4 \TeV$ and  $v=10^{-3}$;
for comparison, the situation where only the LO term is taken into account is also shown (dashed lines).
The  energy spectra  result to be much larger than those obtained in the LO approximation,
by factors $\mathcal{O}(10-100)$.
This is a consequence of having a sizeable $s$-wave annihilation channel opened at 
the next-to-leading order in the $1/r$ expansion.
We have also studied the propagation of these fluxes of stable particles  through the galactic halo
and verified that the enhancement effect under consideration  does not get spoiled by galactic propagation.

\section{Conclusions}
\label{sec:conclusions}

In this talk, I have discussed the relevance of the EW corrections in theories where the  
cross section for DM annihilation into 2-body final states is suppressed.
A gauge-singlet Majorana DM annihilating into two light SM fermions is one such case.

It is found that an efficient lifting of the helicity suppression, opening up a potentially
large $s$-wave, is achieved by including EW bremsstrahlung
and that the resulting energy spectra of stable particles, in the annihilation region, get substantially 
enhanced by this effect by  factors $\mathcal{O}(10-100)$.
This effect crucially affects the predictions for fluxes to be measured at Earth.

These results have a wider generality than the specific model I have considered here.
Reliable computations of energy spectra of stable particles and predictions for their fluxes at Earth -- 
the key observable for DM indirect searches -- cannot prescind from including the effects
of EW radiation.

\section*{Acknowledgments}

This contribution is based on work done in collaboration with 
 P.~Ciafaloni, M.~Cirelli, D.~Comelli, A.~Riotto and A.~Urbano.
I acknowledge support by the Swiss National Science Foundation under
contract 200021-125237.



\end{document}